\documentclass[journal]{vgtc}                

\usepackage{mathptmx}
\usepackage{graphicx}
\usepackage{times}
\usepackage[T1]{fontenc}
\usepackage[latin1]{inputenc}
\usepackage[table]{xcolor}    

\usepackage[normalem]{ulem}

\usepackage[table]{xcolor}
\usepackage{dirtytalk}
\usepackage{xcolor}
\usepackage{soul} 
\definecolor{CX}{rgb}{100, 0, 0}

\usepackage[bookmarks,backref=true,linkcolor=black]{hyperref} 
\hypersetup{
  pdfauthor = {},
  pdftitle = {},
  pdfsubject = {},
  pdfkeywords = {},
  colorlinks=true,
  linkcolor= black,
  citecolor= black,
  pageanchor=true,
  urlcolor = black,
  plainpages = false,
  linktocpage
}

\onlineid{0}

\vgtccategory{Research}

\vgtcinsertpkg

\title{Illusion of Causality in Visualized Data}

\author{Cindy Xiong, Joel Shapiro, Jessica Hullman, and Steven Franconeri}
\authorfooter{
\item
 Cindy Xiong, Jessica Hullman and Steven Franconeri are with Northwestern University. E-mail: cxiong@u.northwestern.edu, jhullman@northwestern.edu, franconeri@northwestern.edu.
\item
 Joel Shapiro is with Northwestern University Kellogg School of Management. E-mail: jshapiro@kellogg.northwestern.edu.

}

\abstract{Students who eat breakfast more frequently tend to have a higher grade point average. From this data, many people might confidently state that a before-school breakfast program would lead to higher grades. This is a reasoning error, because correlation does not necessarily indicate causation -- X and Y can be correlated without one directly causing the other. While this error is pervasive, its prevalence might be amplified or mitigated by the way that the data is presented to a viewer. Across three crowdsourced experiments, we examined whether how simple data relations are presented would mitigate this reasoning error. The first experiment tested examples similar to the breakfast-GPA relation, varying in the plausibility of the causal link. We asked participants to rate their level of agreement that the relation was correlated, which they rated appropriately as high. However, participants also expressed high agreement with a causal interpretation of the data. Levels of support for the causal interpretation were not equally strong across visualization types: causality ratings were highest for text descriptions and bar graphs, but weaker for scatter plots. But is this effect driven by bar graphs aggregating data into two groups or by the visual encoding type? We isolated data aggregation versus visual encoding type and examined their individual effect on perceived causality. Overall, different visualization designs afford different cognitive reasoning affordances across the same data. High levels of data aggregation by graphs tend to be associated with higher perceived causality in data. Participants perceived line and dot visual encodings as more causal than bar encodings. Our results demonstrate how some visualization designs trigger stronger causal links while choosing others can help mitigate unwarranted perceptions of causality.} 

\keywords{Information Visualization, Correlation and Causation, Visualization Design, Reasoning Affordance}

  \teaser{
 \centering
 \includegraphics[width=16 cm]{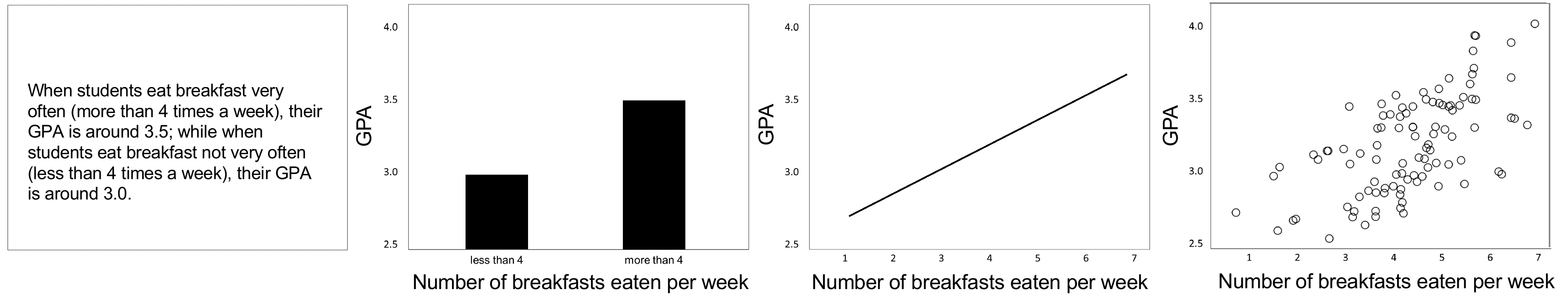}
  \caption{The same data showing the relation between eating breakfast and GPA presented via text, bar graph, line graph or scatter plot. Which depiction makes eating breakfast causing higher GPA seem more plausible to you?}
   \label{teaser}
  }

\CCScatlist{ 
 \CCScat{K.6.1}{Management of Computing and Information Systems}%
{Project and People Management}{Life Cycle};
 \CCScat{K.7.m}{The Computing Profession}{Miscellaneous}{Ethics}
}

\begin{document}


\firstsection{Introduction}

\maketitle

\begin{figure}[htb]
 \centering
 \includegraphics[width=2.5in]{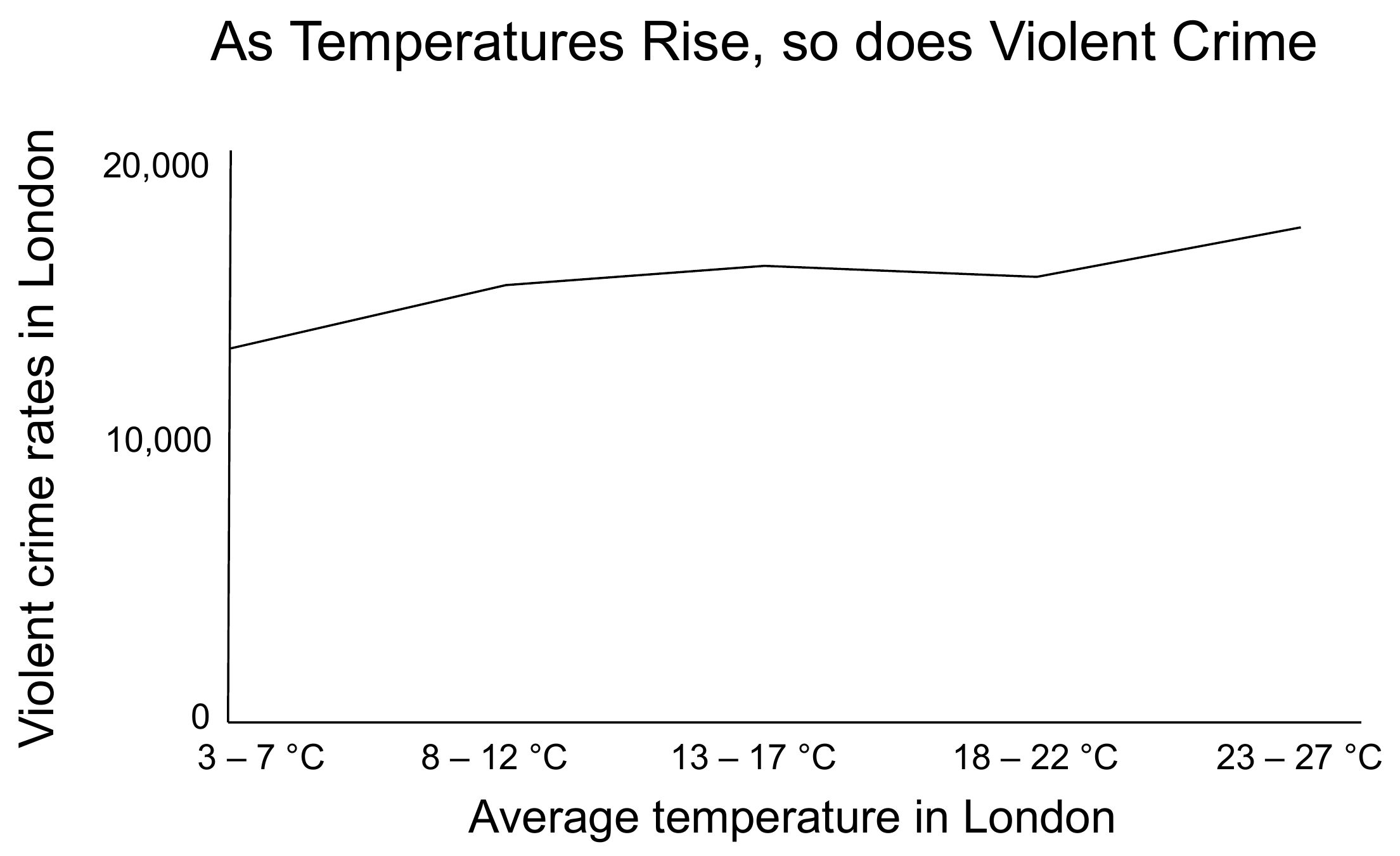}
 \caption{Recreation of BBC news article figure, "Heatwave: Is there more crime in hot weather?"\cite{BBC2019}}
 \label{fig:lineMedia}
\end{figure}

Visualization designs affect decisions. Imagine coming across a piece of BBC news, as shown in Figure \ref{fig:lineMedia}, showing that the number of crimes in London rises with temperature. It can be easy for viewers to conclude that warmer temperature causes violent crimes \cite{BBC2019, matute2015illusions,kahneman2011thinking}.

Concluding causality from the visualized data alone is misguided. We can only establish a correlation - the tendency of two variables changing together - between temperature and crime rate because it is possible that other factors not shown on the graph caused the difference in the number of violent crimes. For example, when the temperature gets warmer, more people go outside, more crimes may happen overall, and thus more violent crimes. If the amount of people outside is kept constant, decreasing temperature would not likely lower crime rates. While the variables illustrated are linked, they are not necessarily causally linked. Yet, people routinely see causal relationships in data.

Confusing correlation with causation is a ubiquitous decision-making error. Just because two factors are correlated (i.e., they tend to co-occur together), it does not mean that one is causing the other. A large portion of work in economics, education, epidemiology, psychology and public health involves analyzing correlations in observed data, which cannot definitively establish causation \cite{robins2000marginal}. Researchers and journalists can sometimes exaggerate causal implications from these results, making it even more difficult for people to decide what kind of conclusions are sound \cite{shiffrin2016drawing, sumner2014association}. This can pave way for misunderstanding of correlation and causation \cite{halpern1998teaching, shaklee1988cause, koslowski1996theory}, potentially having detrimental impact. When researchers or journalists misinterpret or misrepresent correlation for causation, for example, the general public may be misled into thinking correlated factors, such as time of getting vaccinated and time of autism diagnosis, or national debt and GDP growth, are also causally related \cite{dixon2013heightening,reinhart2010growth}.

It is difficult to distinguish causation from correlation~\cite{rothman2012epidemiology}. Even for people who learned 'correlation is not causation' with classroom examples, it could still be challenging to apply their learning to new contexts \cite{shtulman2012scientific, rhodes2014explaining}. Because establishing causal inference is complex, even trained scientists can sometimes struggle with correlation and causation \cite{halpern1998teaching}. We are interested in whether a simple change in the visualization design can reduce unwarranted conclusions of causality.

Although many have looked at the effect of visualization designs on \textit{perceptual} analytic tasks such as determining anomalies or estimating data trends \cite{saket2018task, correll2014error,croxton1927bar,eells1926relative,harrison2014ranking, spence1991displaying,cleveland1984graphical,kay2016ish}, researchers have only begun to explore the effect of visualization design on \textit{cognitive reasoning} tasks, such as understanding uncertainty \cite{hullman2018imagining,kale2019hypothetical}, persuading attitude or belief change \cite{kim2019,pandey2014persuasive} or eliciting empathy \cite{boyEmpathy}. Previous work has demonstrated visualization designs could influence data interpretation. For example, many people conclude ``on average, Dutch are taller than Americans'' from a bar graph visualizing the height of Americans and Dutch, but when the same information is visualized with a line graph, people are more likely to conclude ``people get taller as they become more Dutch.''\cite{zacks1999bars}. We suspect visualization designs can also afford different cognitive reasoning routines in data, triggering perceived causal links more or less strongly in data. \newline
\textbf{Contribution:} We contribute three empirical studies to examine how visualization design can afford different interpretations of correlation and causation in data. Experiment 1 finds that higher proportion of people draw causal conclusions from bar graphs and plain text compared to scatter plots and line graphs. Experiment 2 and 3 found this effect to be driven by data aggregation as well as visual encoding marks. Less aggregation (binning data into more groups) and dot encoding marks (instead of rectangular bars and lines) reduced the strength of perceived causal links in data. This work provides a first step towards design guidelines that facilitate better interpretations of correlation and causation in data.

\begin{table*}[htp]
\rowcolors{2}{gray!25}{white}
    \scriptsize
    \caption{Correlation and causation plausibility ratings for the four selected statement sets from the pilot experiment.}
    \begin{tabular}{lp{10.4cm}lc}
  Variables & Statement & Type & Plausibility Rating\\
   \hline

    spending and fitness & People who spend more on admission to sporting events tend to be more physically fit. & correlation &  65.91\\
  &  If people were to spend more on admission to sporting events, they would be more fit. & causation & 52.52\\     
  
    smoking and cancer & People who smoke more have a higher risk of getting lung cancer.& correlation & 88.14\\
    &  If people smoke more, they would have higher risk of getting lung cancer.& causation & 91.19\\
      
 breakfast and GPA &  Students who more often eat breakfast tend to have higher GPA. & correlation & 83.86\\
   &  If students were to eat breakfast more often, they would have higher GPA. & causation & 78.43\\

   internet and homicide & When there are more people using Internet Explorer, the homicide rates in the United States tend to be higher. & correlation & 35.57\\
   &  If more people used Internet Explorer, there would be more homicide in the United States.& causation & 28.38\\
   
\hline
\end{tabular}
\label{pilot_stories}
\end{table*}

\section{Related Work}
Visualization design can influence the \textit{type} of information extracted and the inferences made from data. In perceptual analytic tasks, choosing the appropriate visualization designs can improve the accuracy and efficiency of information extraction. Spatially upward trends are conventionally associated with increasing values, even when the axes are reverse labelled \cite{padilla2018case}. Bar graphs facilitate finding clusters, line graphs facilitate finding correlations and scatter plots facilitate finding outliers \cite{saket2018task,zacks1999bars}. Visual marks, such as rectangular bars, lines or dots, can support different inferences about data relations based on their geometric properties. For example, lines indicate connection, arrows indicate dynamic (or causal) information \cite{heiser2006arrows}, and scattered dots each represents a value of an individual subject or collection \cite{friendly2005early}.

In higher-level decision tasks, visualization design also influences data interpretation and decision making \cite{dimara:2017:Attraction, cho2017anchoring}. People interpret climate data differently depending on whether the visualization presented percentile information versus showing the range \cite{daron2015interpreting}. In bar graphs depicting average values, people judge data values that fall within the bar as being more likely to be part of the data set than points outside the bar, despite being equidistant from the mean \cite{newman2012bar,correll2014error}. People can be more easily persuaded by tabular designs of data when they hold strong attitudes against the depicted topic, but more easily persuaded by bar graphs when they have no strong attitudes \cite{pandey2014persuasive}. People also rely on visual salience of attributes to interpret data \cite{jarvenpaa1990graphic}. These examples support that different visualization designs of the same data could afford different interpretation of data at a higher-level, which may extend to causal or correlational interpretations. 

\begin{figure}[hb]
 \centering
 \includegraphics[width=3.35in]{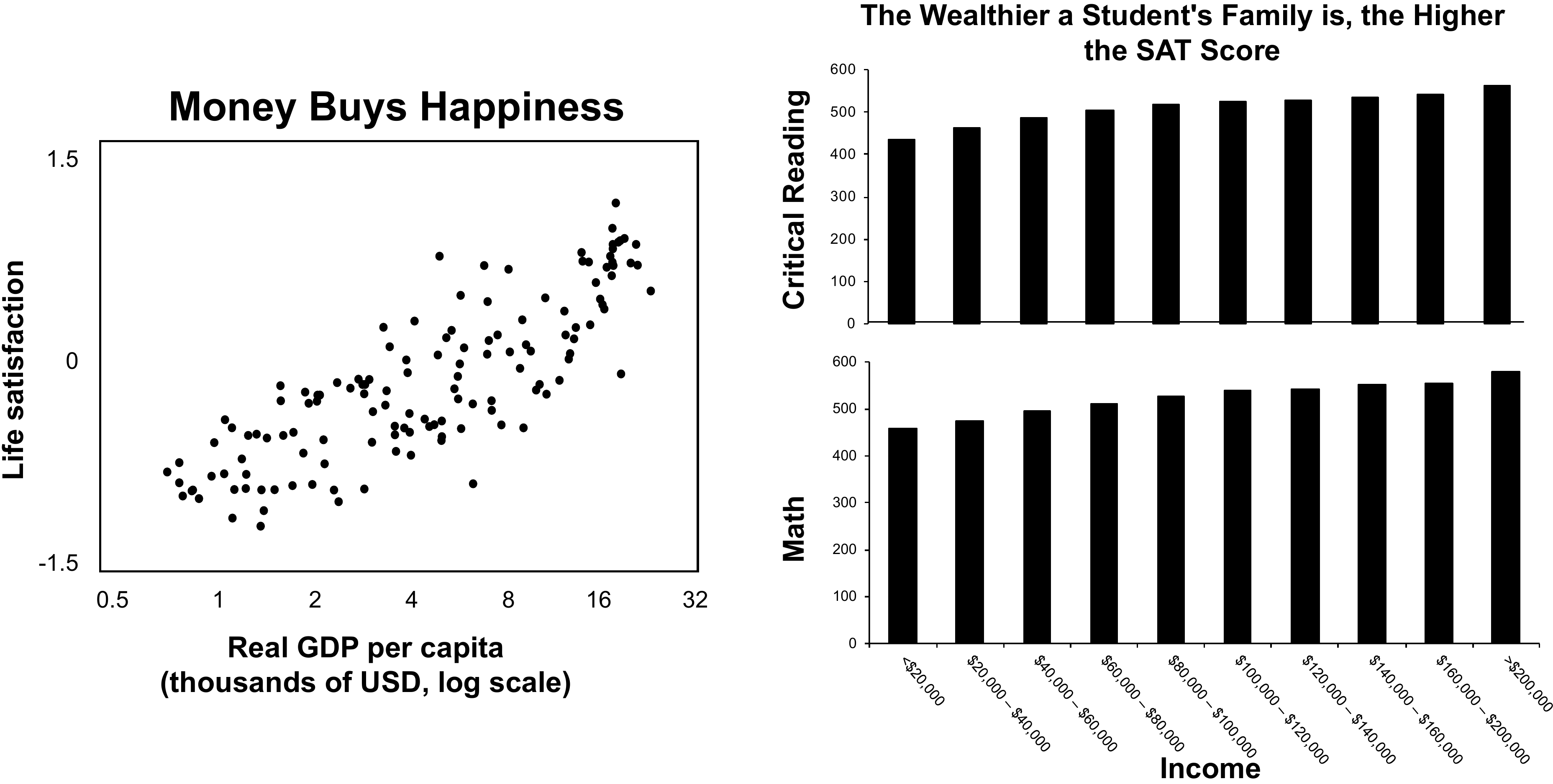}
 \caption{Left: recreation of NPR article "Money Buys Happiness," showing a correlation between GDP and life satisfaction \cite{npr2011}. Right: recreation of the Washington Post news article figure, "Researchers have debunked one of our most basic assumptions about how the world works," showing a correlation, but not causation, between income and SAT scores\cite{wp2016}.}
 \label{fig:barMedia}
\end{figure}

What types of visual formats are commonly used to present correlated data? Bar graphs, line graphs and scatter plots are common ways to depict correlated data in media \cite{BBC2019,wp2016}, alongside text, as shown in Figure \ref{fig:lineMedia} and Figure \ref{fig:barMedia}. We investigate how bar graphs, line graphs, scatter plots and text influence causal reasoning of data.

Research on perceptions of causality indicates that they can be context-dependent, in addition to being visualization design-dependent. When the evidence presented aligns with people's prior experience, emotional response or beliefs, they become more likely to judge the evidence as sound \cite{shah2017makes}. People often perceive high causality when they judge the evidence as sound and stop thinking through other possible explanations \cite{kahneman2011thinking}. Prior work suggests that persuasiveness of visualized data depends on both context (does the topic align with the viewers' prior beliefs?) and visualization designs (tabular design or bar graphs) \cite{kim2017,kim2019,pandey2014persuasive}. Thus we also examine the effect of context by testing a set of paired variables that vary in the plausibility of their causal link, which we establish in a pilot experiment.

The task people perform when viewing the visualizations may also influence the conclusions they draw. Our experiments consider two common tasks people perform when interacting with data. The first is a \textit{judgment} task in which they decide whether they agree or disagree with the presented information. For example, media often present people with visualizations alongside text describing a correlational or a causal relation between depicted variables \cite{bromme2014public}. In this scenario, information consumers have to decide how much they agree with the description based on the visualized data. Judgment tasks can be evaluated by comparing participant ratings of how much they agree with statement describing a correlation or a causation. The second is a \textit{generative} task where people have to independently interpret a visualization to draw their own conclusions. One example is when a data analyst working to make sense of their data hoping to deliver a research report on the newest scientific findings. In this scenario, the data analyst has to actively interpret some visualizations and generate a conclusion. Generative tasks may shed more insights on how participants interpreted data and arrived at possible correlational/causal conclusions, but because they are open-ended, they tend to be more difficult to formally evaluate. In our pilot experiment, we asked participants to generate interpretations of data, then used their interpretations to develop a taxonomy to facilitate analysis of generative tasks in Experiment 1.

\section{Pilot Experiment}
Taking inspiration from the anecdotes of a set of local instructors of research methods and data analytics, we generated 19 potential variable pairs, from those with plausible causal relations to those with implausible causal relations. We conducted a pilot experiment to test the perceived correlation and causation of these variable pairs. 

Specifically, we surveyed 21 participants for their perceived plausibility of correlational and causal relations of the 19 variable pairs through Qualtrics on Amazon's Mechanical Turk (MTurk) \cite{qualtrics2013qualtrics}. Participants viewed the 19 correlation and causation statement sets in random orders. For each pair, they first interpreted its message and justified their reasoning in a text box. This is the \textit{generative} task. Then, on a separate page, they read a correlation statement and a causation statement, as shown in Table \ref{pilot_stories}. The correlation statement accurately describes the relation between the depicted data variables, while the causation statement attributes causal relations to the depicted data variables. They gave a plausibility rating for each (0 = extremely implausible, 100 = extremely plausible).  This task reflects the \textit{judgment} tasks people would perform in real life.

\subsection{Picking Statements}
\begin{figure}[htb]
 \centering
 \includegraphics[width=3.2in]{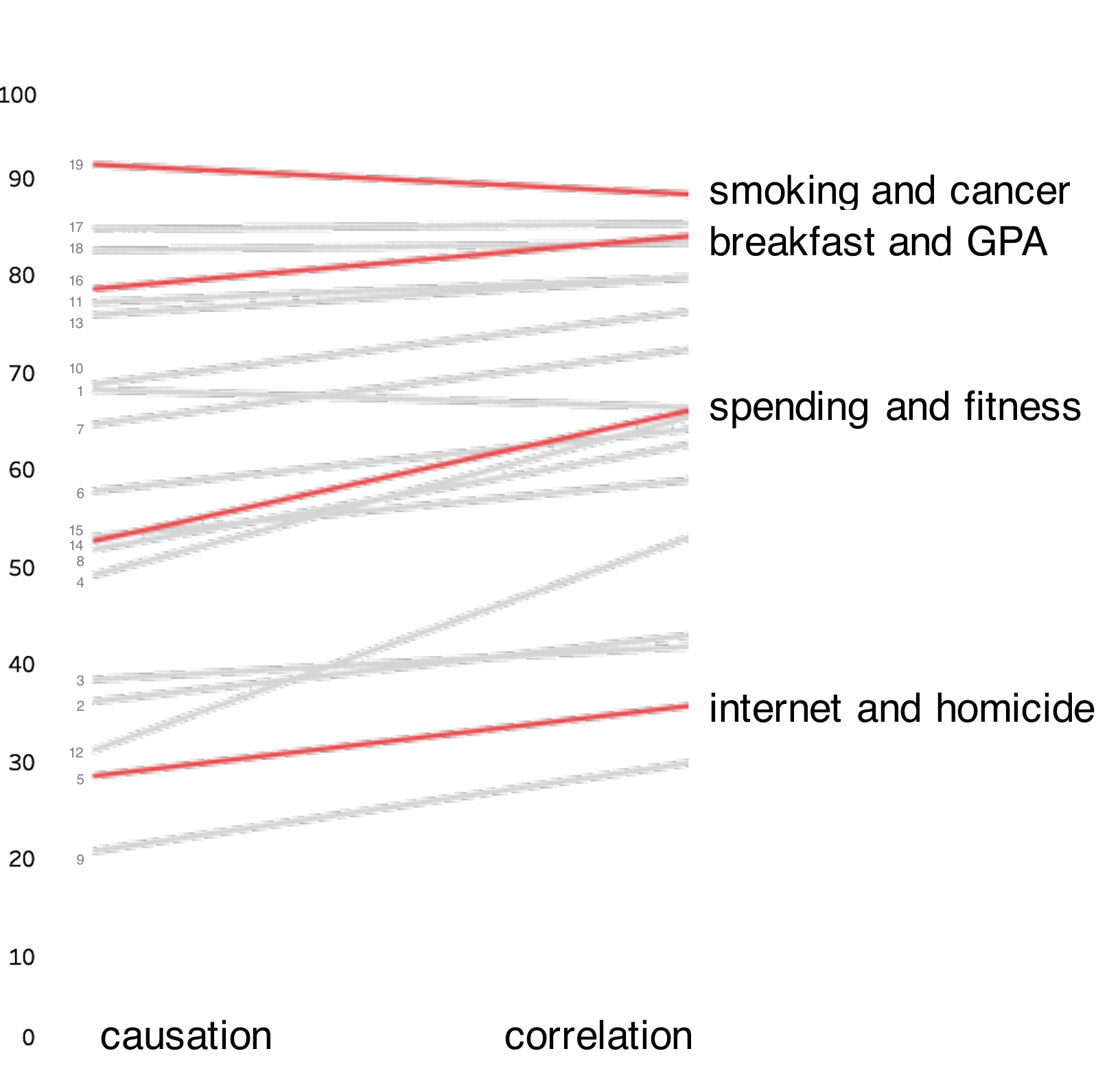}
  \caption{Pilot results. Grey numbers indicate the index of the 19 statements, details see supplementary. The line positions represent mean correlation and causation plausibility ratings. Red lines are the correlation and causation plausibility ratings for the selected contexts, intended to cover a range of plausibility. 
  }
 \label{pilot-plausibility}
\end{figure}

The participants rated their perceived plausibility of both the correlation and causation statements. Table \ref{pilot_stories} shows the four contexts we picked with varying plausibility. These four context differed significantly in their perceived correlation and causation ratings, based on an analysis of variances, as shown in Figure \ref{pilot-plausibility}. We visualized information using these four contexts in Experiment 1 to investigate the effect of visualization design on perceived causality.

\subsection{Qualitative Coding: Interpretation Taxonomy}
To provide a structured way of interpreting participants' statements in our experiments, we analyzed the freeform written response from the generative task in the pilot, in which participants drew conclusions from the information and justified their correlation and causation ratings, to create a taxonomy to characterize these conclusions in the experiment. We identified six dimensions that could help us characterize and evaluate the conclusions participants generated -- whether the participant concluded correlation, concluded causation, mentioned third variables, grouped variables together, made direct observations or explicitly stated the data to be inconclusive. Each response is coded independently on these six dimensions, which means the same response could fit into multiple categories.

\textbf{Distinguishing Correlation from Causation: }Referencing past work outlining a taxonomy of causal arguments \cite{oestermeier2000verbal}, we looked for causal inference patterns in the verbal responses in the generative task, to distinguish a causal interpretation from a correlational one. Specifically, words such as "causes", "leads to" and "results in" depending on the context, suggests causal interpretations, while phrases such as "as X increases, Y tend to increase" were classified as correlational interpretations. 

\textbf{Mentioning Third Variables: }If participants discussed variables not depicted in the visualization as influencing the relations between the two depicted variables, we additionally labelled the response as "considered third variables." 

\textbf{Grouping Variables: }Participants could also group the levels of a variable together when justifying their reasoning. For example, one could say "when X is high, Y is high, but when X is low, Y is low," which arbitrarily divides the x--variable into two dimensions. 
Grouping of variables may be associated with misattributed causal relations. Thus we examine variable-grouping as part of our taxonomy.

\textbf{Direct Observations: }We also anticipated that not all participants would provide high-level reasoning. Some could make direct observations, stating the values depicted in a visualization verbatim. "When X is 2, Y is 3" and "there is a vertical line starting at 15000" are both instances of direct observations. 

\textbf{Inconclusive Responses: }Participants could also deem the amount of data present inconclusive without drawing any correlational or causal conclusions.

\section{Experiment 1 Causality in Context}
\begin{figure}[htb]
 \centering
 \includegraphics[width=3.45in]{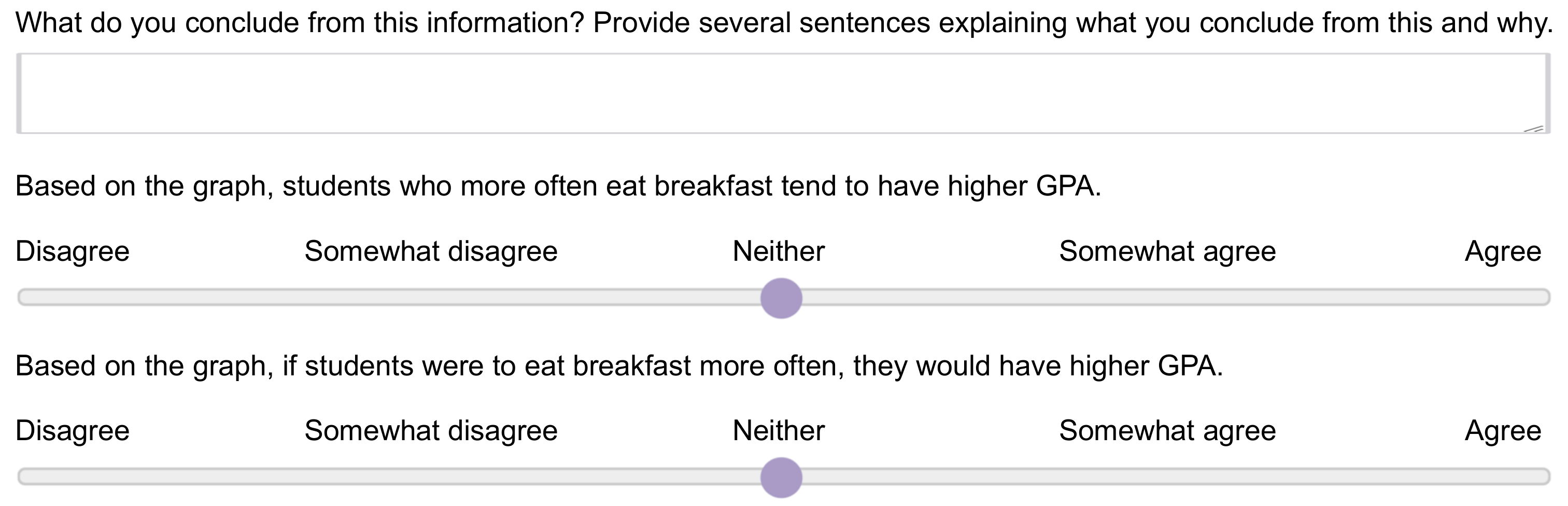}
  \caption{Example of generative task (top) and judgment task (middle and bottom) in Experiment 1. The three questions were shown on \textit{separate} pages in Qualtrics in the order from top to bottom.}
 \label{judgmentgenerativeexample}
\end{figure}

Experiment 1 investigates whether visualization design influences how people interpret correlation and causation in data, using the four variable pairs selected from the pilot experiment. We asked participants to complete both judgment and generative tasks, in which they rate how much they agree with a correlation or causation statement, and verbally interpret the information and justify their judgment task reasoning, as shown in Figure \ref{judgmentgenerativeexample}. 

\subsection{Participants}
Participants were recruited through the Human Intelligence Task (HIT) postings on MTurk. We excluded workers who are not based in the United States, have an approval rate below 95\%, failed the attention checks, entered nonsensical answers for the free response questions or failed the graph reading comprehension checks (details of these checks are included in the supplementary materials). An omnibus power analysis based on pilot effect sizes suggested a target sample of 136 participants would give us 95\% power to detect an overall difference between visualization designs at alpha level of 0.05. We iteratively surveyed and excluded participants until we reached this sample size. 

\subsection{Design}
\begin{figure}[htb]
 \centering
 \includegraphics[width=3.1in]{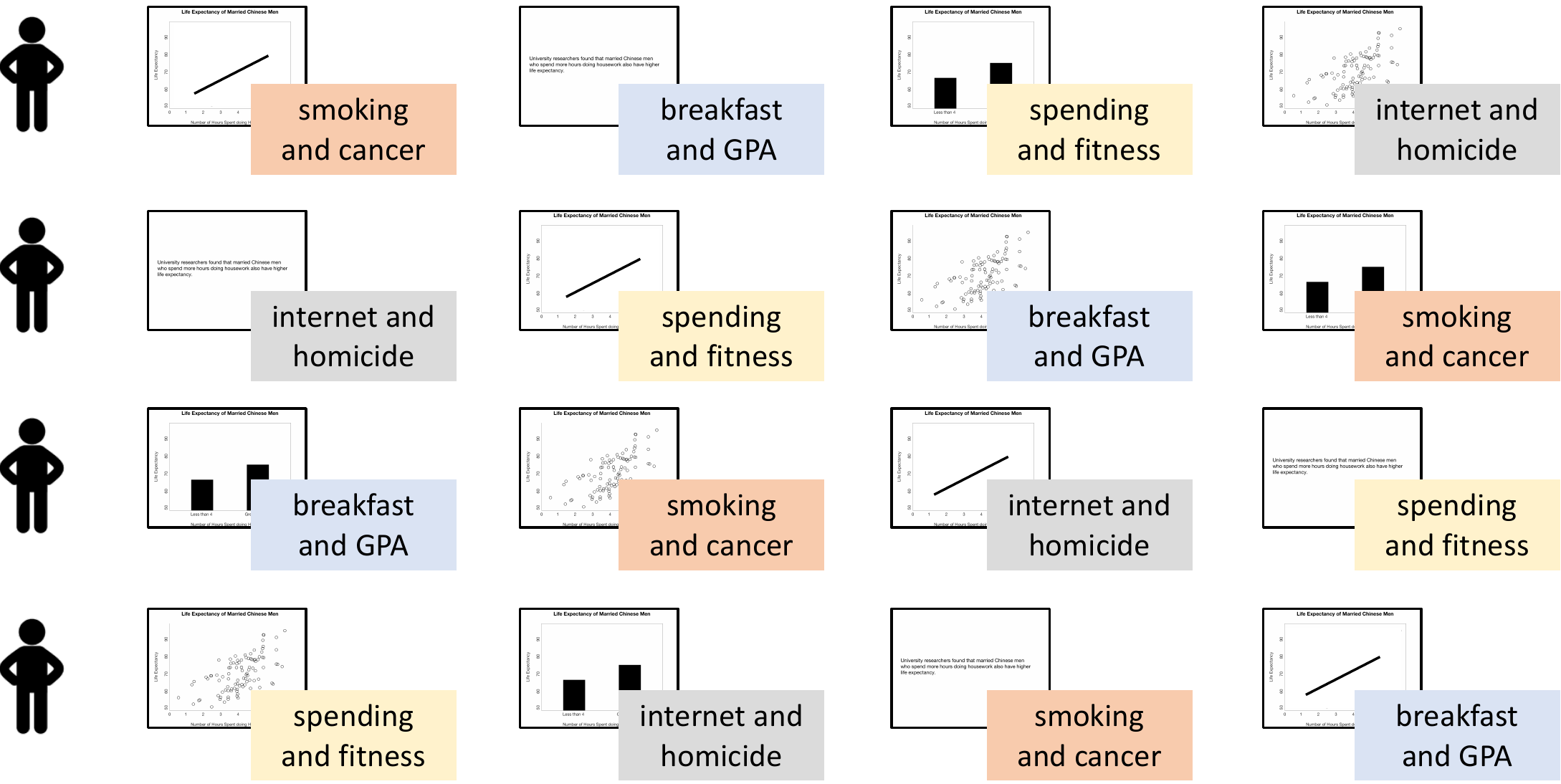}
 \caption{Graeco-Latin Square design showing the four conditions for Experiment 1. Each row represents a condition. Each column represents the order in which the participants saw the stimuli, with the left-most seen first and the right-most seen last.}
 \label{exp1_latinSquare}
\end{figure}

This experiment had a $4\times4$ Graeco Latin Square design. As shown in Figure \ref{exp1_latinSquare}, each participant saw four sets of data in the four variable pairing chosen from the pilot experiment, presented using four visualization designs. We will refer to the variable pairing as `\textit{context.}' We replicated each condition 34 times with different participants to increase the reliability in our measures. We chose three simple visualization designs commonly seen in media and education \cite{BBC2019,wp2016,tufte2001visual,knaflic2015storytelling} -- bar graphs, line graphs and scatter plots as well as a plain text, as shown in Figure \ref{teaser}. The plain text was written to parallel the bar graph, including identical information in which one variable (X) was arbitrarily divided into two groups and the corresponding average value for the other variable (Y) at those two groups were specified.

Our independent variables are the visualization design and context plausibility. Visualization design is a categorical variable indicating the design we presented the information to the participants, which could be bar graphs, line graphs, scatter plots or plain text. Context plausibility is the correlation and causation statement plausibility collected from the pilot experiment, which is a continuous variable from 0, extremely implausible, to 100, extremely plausible. We recorded the order in which the participants viewed the visualizations. We also collected demographic information such as participant age, gender, political orientation and level of education.

There were two dependent variables. Four researchers blind to both the study design and the condition manipulations coded the response in the \textit{generative} task based on the interpretive taxonomy, and the participant count in each category (e.g., direct observation) was one dependent variable. The other dependent variable was participants' ratings on how much they agreed with the correlation and causation statements listed in Table \ref{pilot_stories} in the \textit{judgment} task.

\subsection{Materials}
\begin{figure}[htb]
 \centering
 \includegraphics[width=3in]{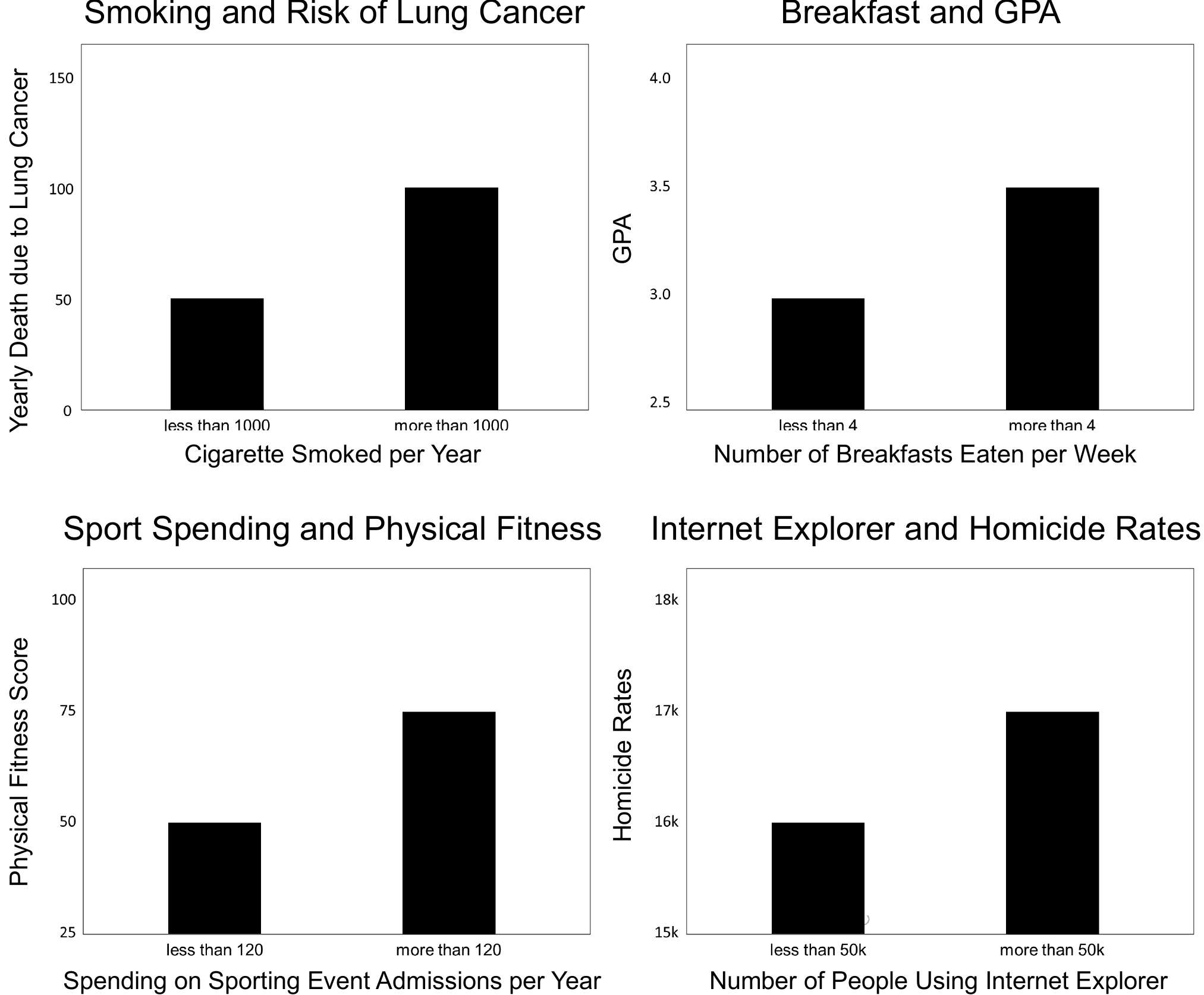}
 \caption{The bar graph stimulus in the four contexts.}
 \label{bargraphs_difflabels}
\end{figure}

We used MATLAB to randomly generate 100 pairs of data points from a normal distribution with a correlation of 0.6 to avoid ceiling and floor effect of rating the underlying correlation as too high or too low. We visualized this dataset into a bar graph, line graph and scatter plot, as shown in Figure \ref{teaser}. To ensure all participants viewed the same visualized data across all conditions, we relabeled the axis to fit the context without changing the underlying dataset. For example, Figure \ref{bargraphs_difflabels} shows the bar graph depicted in the four contexts.

\subsection{Procedure}
Upon accepting the HIT, participants clicked on a Qualtrics link to access the experiment. Participants completed the four task trials and finished with demographic questions. On each trial, participants viewed a visualization (bar, line, scatter or text) and answered two graph reading comprehension check questions. They then completed the generative task in which they wrote several sentences explaining what they concluded from the visualization and why. This was followed by the judgment task in which participants read a correlation and a causation statement (presented separately on two pages), and rated how much they agree with each on a scale from 0 (disagree) to 100 (agree), as shown in Figure \ref{judgmentgenerativeexample}.

\begin{figure*}[ht]
 \includegraphics[width=17.2cm]{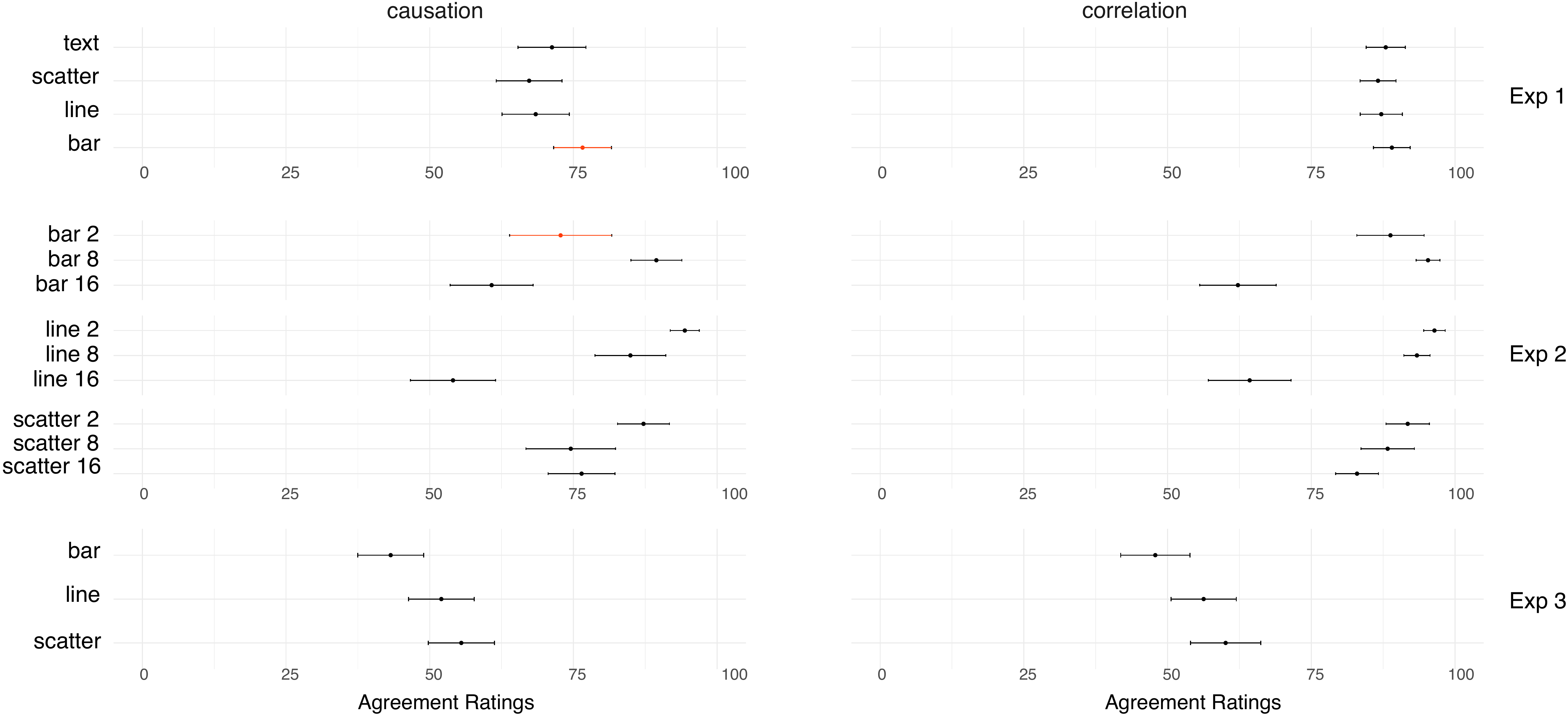}
 \caption{Quantitative results from all three experiments showing participants' correlation and causation agreement ratings.}
 \label{fig:allResults}
\end{figure*}

\subsection{Causation Judgment Results}
We used a mixed-effect linear model to fit the causation ratings \cite{bates2005fitting}, which was how much each participant agreed with the causation statements, under the four visualization designs (bar, line, scatter and text). For fixed effects, we used visualization design, causation statement plausibility, trial order and demographic information (age, gender, education and political orientation) as predictors. Because it seemed plausible that certain combinations of contexts (pairs) and visualization designs could interact to increase or lessen perceived causality (i.e., based on conventions for showing data in certain domains), we also considered an interaction between visualization design and causation statement plausibility. We used a random intercept term accounting for individual differences as random effects.

The regression model indicated a relatively large effect of causation statement plausibility (context), $\chi^2$=162.70, $\eta_{partial}^{2}$=0.274,$p$<0.001, a relatively small effect of visualization design ($\chi^2$=11.65,$\eta_{partial}^{2}$=0.026,$p$<0.01), and negligible interaction effect between causation statement plausibility (context) and visualization design ($\chi^2$=0.97,$\eta_{partial}^{2}$=0.002,$p$=0.81). Referencing Figure \ref{fig:allResults}, participants rated bar graphs to be the most causal ($M$=76.59, $CI_{95\%}$=[71.51, 81.76]) and text the second most causal ($M$=71.26, $CI_{95\%}$=[65.30, 77.23]). This largely agreed with the results from the generative tasks where participants also made causal interpretations and the most group-wise comparisons in bar graphs and text. Given the similarity between bar graphs and text, which was written to contain identical information as the bar graph (grouping the data into two groups), we suspected that perceived causality differed between visualization designs because information was organized and presented differently among them. 

Line graphs and scatter plots, unlike bar graphs and text, did not group variables together. Participants rated line graphs ($M$=68.43, $CI_{95\%}$=[62.52, 74.35]) and scatter plots ($M$=67.29, $CI_{95\%}$=[61.52, 73.07]) the least causal, which were the two designs with the most correlation interpretation in the generative task. This suggests that the effect of visualization design on perceived causality could be driven by data aggregation and visual encoding marks.

There is negligible effect of the order the visualizations were presented ($\chi^2$=0.11,$\eta_{partial}^{2}$=0.002, $p$=0.74), which means perceived causation does not depend on what was presented to them previously nor was there a learning effect. 
Results also indicated a comparatively small effect of gender ($\chi^2$=4.23,$\eta_{partial}^{2}$=0.007,$p$=0.040), such that male participants gave higher causation ratings, and education ($\chi^2$=0.4.53, $\eta_{partial}^{2}$=0.011,$p$=0.033), such that participants with higher levels of educating gave lower causation ratings.

\subsection{Correlation Judgment Results}
We used a similar mixed-effect linear model to predict how much each participant agreed with the correlation statements. We kept all predictors the same with the exception of swapping the causation statement plausibility with the correlation statement plausibility. Only correlation statement plausibility had a sizable effect predicting perceived correlations ($\chi^2$=71.02,$\eta_{partial}^{2}$=0.141,$p$<0.001), there was negligible effect of visualization design ($\chi^2$=1.98,$\eta_{partial}^{2}$=0.005,$p$=0.58), a small interaction between the two ($\chi^2$=6.15,$\eta_{partial}^{2}$=0.012,$p$=0.10), a tiny effect of education ($\chi^2$=2.99,$\eta_{partial}^{2}$=0.007,$p$=0.08), such that participants with higher levels of education gave lower correlation ratings. There were negligible effects of order, age and gender (details included in the supplementary materials). We can see this from the similar correlation confidence intervals in Figure \ref{fig:allResults}. This suggests visualization design does not significantly influence people's judgment of correlation from data, at least when participants were given a concrete context.

\begin{figure*}[htb]
 \centering
 \includegraphics[width=17.8cm]{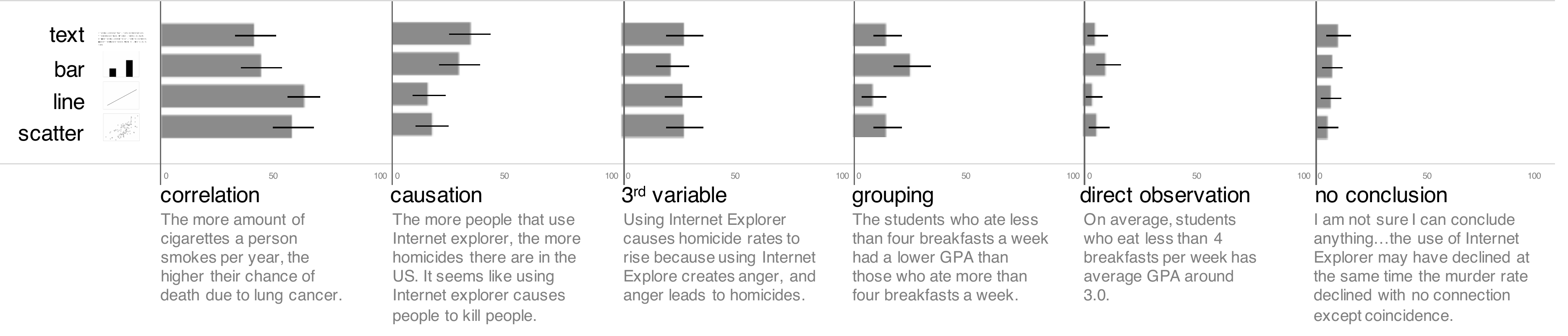}
 \caption{Qualitative coding results of Experiment 1. Each bar represents the percentage of participants that mentioned the indicated dimension (e.g., third variable) for a certain visualization design.}
 \label{qual_exp1_exp2}
\end{figure*}

\subsection{Qualitative Results from Generative Task}
Each of generative task responses was coded as "yes" or "no" on each of the six categories, as shown in the top row of Figure \ref{qual_exp1_exp2}.

\textbf{Correlation Conclusions:} Many participants appropriately inferred correlation between depicted variables, using words and phrases such as "tend to" and "the more X the more Y." A chi-square test of independence with Bonferroni adjustment suggests that varying proportion of participants drew correlation conclusions from different visualization designs ($\chi^2$=27.84, $p$<0.001). On average, in 75.7\% of the trials participants drew correlation conclusion from line graphs ($CI_{95\%}$=[68.7, 82.9]), 69.1\% from scatterplots ($CI_{95\%}$=[61.4, 76.9]), 52.9\% from bar graphs ($CI_{95\%}$=[44.6, 61.3]), and 50.0\% from text ($CI_{95\%}$=[41.6, 58.4]). Figure \ref{qual_exp1_exp2} shows one example of a correlation interpretation.
 
\textbf{Causal Conclusions} Among the participants who generated causal conclusions from the data, some used causation suggestive words such as ``leads to'' or ``causes'', while others seemed to have assumed causation without using causation suggestive words. Some of these participants dismissed the visualized information as illogical because the causal relation they interpreted went against their belief or intuition. As a result, some did not reach a conclusion from the visualization, not because they were aware that correlation is not causation, but because they thought the visualization was depicting a causal relation that did not make sense to them. 

For example, in response to the "spending and fitness" visualization, one participant suggested that the visualization did not make sense because "there is no correlation between the two," mistaking correlation for causation. In this case, the participant seemed to understand the notion that correlation is not causation, but assumed that the visual results implied more than just correlation nonetheless. We coded the response as both "causation" and "no conclusion." 

\label{barExperiments}
There were also two participants who mentioned "experiments" in their responses with bar graphs, even though we specifically noted that the visualizations are generated from survey data. It is possible that some people associate bar graphs with controlled experiments, from which causal conclusions can be validly drawn.

We found several common characteristics among participants who did not assume causal relations. They questioned the directionality and predispositions, or mentioned third variables at play. For example, in the "breakfast and GPA" context, participants who did not assume causation questioned whether it is people who ate breakfast more were more likely to get good grades, or that people who were more likely to get good grades were more organized, and thus more likely to get up early and eat breakfast.

A chi-square test of independence revealed an overall effect of visualization design on whether people drew causal conclusions as defined by their generated responses ($\chi^{2}$=21.77, $p$<0.0001). As shown in the causation column in Figure \ref{qual_exp1_exp2}, in 39.0\% of the trials participants drew causal conclusion from text ($CI_{95\%}$=[30.8, 47.2]), in 33.8\% from bar graphs ($CI_{95\%}$=[25.9, 41.8]), in 20.6\% from scatter plots ($CI_{95\%}$=[13.8, 27.4]), and 18.4\% from line graphs ($CI_{95\%}$=[11.9, 24.9]).

\textbf{Third Variables} Visualization designs might influence whether people think of third variables when drawing conclusions from visualizations. We observed participants justifying both correlation and causation by connecting a third variable to the two visualized. For example, in the "internet and homicide" context, one participant speculated that "\textit{using Internet Explorer causes homicide rates to rise because using Internet Explore[r] creates anger, and anger leads to homicides.}" Anger is not visualized on the graph, therefore it is a third variable.

A chi-square test of independence suggested that there was \textit{no} relation between visualization design and mentioning of third variables ($\chi^{2}$=2.03, $p$=0.57), suggesting no particular visualization design makes people more or less likely to think of third variables, as shown in the $3^{rd}$ variable column in Figure \ref{qual_exp1_exp2}. On average, in 30.9\% of the trials participants mentioned third variables in scatter plots ($CI_{95\%}$=[23.1, 38.7]), 30.9\% in text ($CI_{95\%}$=[23.1, 38.7]), 30.2\% in line graphs ($CI_{95\%}$=[22.4, 37.9]), and 24.3\% in bar graphs ($CI_{95\%}$=[17.1, 31.5]).

\textbf{Grouping in Response} We observed an overall effect of visualization design on the number of group-wise comparisons made ($\chi^{2}$=15.57, $p$<0.001). Researchers coded responses as group-wise comparisons when the participant described the visualized data in two groups by one dimension and compared the two grouped values in the other dimension. For example,
\vspace{1.5pt}\newline
\noindent\textit{"The students who ate less than four breakfasts a week had a lower GPA than those who ate more than four breakfasts a week."}

In 27.9\% of the trials participants made group-wise comparisons of variables in bar graphs ($CI_{95\%}$=[20.4, 35.5]), 16.2\% in text ($CI_{95\%}$=[9.99, 22.4]), 16.2\% in scatter plots ($CI_{95\%}$=[9.99, 22.4]), and 9.6\% in line graphs ($CI_{95\%}$=[4.6, 14.5]).

\textbf{Direct Observations} While no visualization elicited more direct observations than others $\chi^{2}$=5.09, $p$=0.17), we observed several direct, number-specific comparisons instead of global pattern or trend observations across all designs. For example, when viewing a bar visualization on "breakfast and GPA," one participant concluded --
\vspace{1.5pt}\newline
\noindent\textit{"On average, students who eat less than 4 breakfasts per week has average GPA around 3.0."}

As shown in Figure \ref{qual_exp1_exp2}, in 11.0\% of the trials participants made direct observations in bar graphs ($CI_{95\%}$=[5.8, 16.3]), 6.6\% in scatter plots ($CI_{95\%}$=[2.4, 10.8]), 5.9\% in text ($CI_{95\%}$=[1.9, 9.8]), and 4.4\% line graphs ($CI_{95\%}$=[0.96, 7.9]). 

\textbf{No Conclusions} All visualizations elicited the same proportion of non conclusions ($\chi^{2}$=2.57, $p$=0.46). In 11.0\% of the trials participants drew no conclusion in text ($CI_{95\%}$=[5.8, 16.3]), 8.1\% in bar graphs ($CI_{95\%}$=[3.5, 12.7]), 7.4\% in line graphs ($CI_{95\%}$=[3.0, 11.7]), and 5.9\% in scatter plots ($CI_{95\%}$=[1.9, 9.8]). 

We observed two types of no conclusion responses, one in which participants inferred causality from the visualization but decided to draw no conclusion because it went against their intuition, and the other in which participants made a conscious decision not to. This could be a result of them choosing to be skeptical about the completeness of the information or being aware of "correlation is not causation." For example, in response to the "internet and homicide" context, one participant wrote\vspace{1.5pt}\newline
\noindent\textit{"I am not sure I can conclude anything ---the use of Internet Explorer may have declined at the same time the murder rate declined with no connection except coincidence."}
\vspace{3pt}

In general, many people drew from their personal experience or knowledge to make sense of the visualized information. Congruent with prior research, most participants' first intuition is to justify a potential relation between the variables visualized, despite the plausibility of the causal link \cite{ibrahim2016using, kahneman2011thinking}. Few stopped and thought of "counter examples," questioned the validity of the data, or showed clear signs of understanding that correlation is not causation. 

Some participants used "template" words or phrases, such as "correlation is not causation" or "Y tend to increase with varying levels of X" to frame their conclusions. For example, one participant made the following conclusion in the "internet and homicide" scenario.
\vspace{1.5pt}\newline
\noindent\textit{"The graph shows that in cities with more people using Internet Explorer, there tend to be many more homicides. While the results are pretty clear, I think "correlation is not causation" should be applied here. I'm not a scientist, but I don't think the two variables are really related in any meaningful way."}

It is also apparent when a participant only memorized the phrase "correlation is not causation" without truly understanding the concept. They read correlation from the data, and assumed the data to be telling a causal story as they confuse correlation for causation. But, because they were superficially aware that "correlation is not causation," they dismissed the \textit{correlation} in data despite the observable correlation in data. For example, this participant was clearly aware of the phrase "correlation is not causation," but instead of critically thinking through third variables or other possibilities, quickly dismissed the data and the apparent correlation. \vspace{1.5pt}\newline
\noindent\textit{"With only this information I can't conclude anything since I do not see any correlation. In my opinion these two variables are uncorrelated..."}

Furthermore, all participants interpreted the visualization assuming the X -> Y directionality, such as "as X increases Y increases." For people who made causal conclusions, all of them described the x-axis variable as the cause and the y-axis variable as the effect. This suggests that there may exist a conventional interpretation of causality in data for the x-axis variable to be seen as the cause and the y-axis variable to be seen as the cause.

\subsection{Discussion of Experiment 1} 
In general, the quantitative and qualitative results told similar stories of how, when given specific pairs of common variables, people perceived causality as more likely in bar graphs and less likely in scatter graphs. Context also had a relatively large effect on perceived causality, but the effect of visualization design on perceived causality was not context dependent. We took away the specific pairs of common variables in subsequent experiments to further examine \textit{how} visualization designs influence perceived causality.

\section{Experiment 2 Aggregation levels}
Experiment 1 found that people perceived high causality from bar graphs and low causality from scatter plots. But is this driven by properties of the visual encoding marks (e.g., rectangular bars versus circular points versus lines), or by how aggregated data is? For example, the bar graph we showed aggregated the data into 2 groups while the scatter plot did not aggregate any data, showing each data point individually. 
Experiment 2 tested the effect of the amount of aggregation in data on perceived causality, and whether the visual encoding marks interact with this effect by comparing bar graphs, line graphs and scatter plots.

\begin{figure}[htb]
 \centering
 \includegraphics[width=8cm]{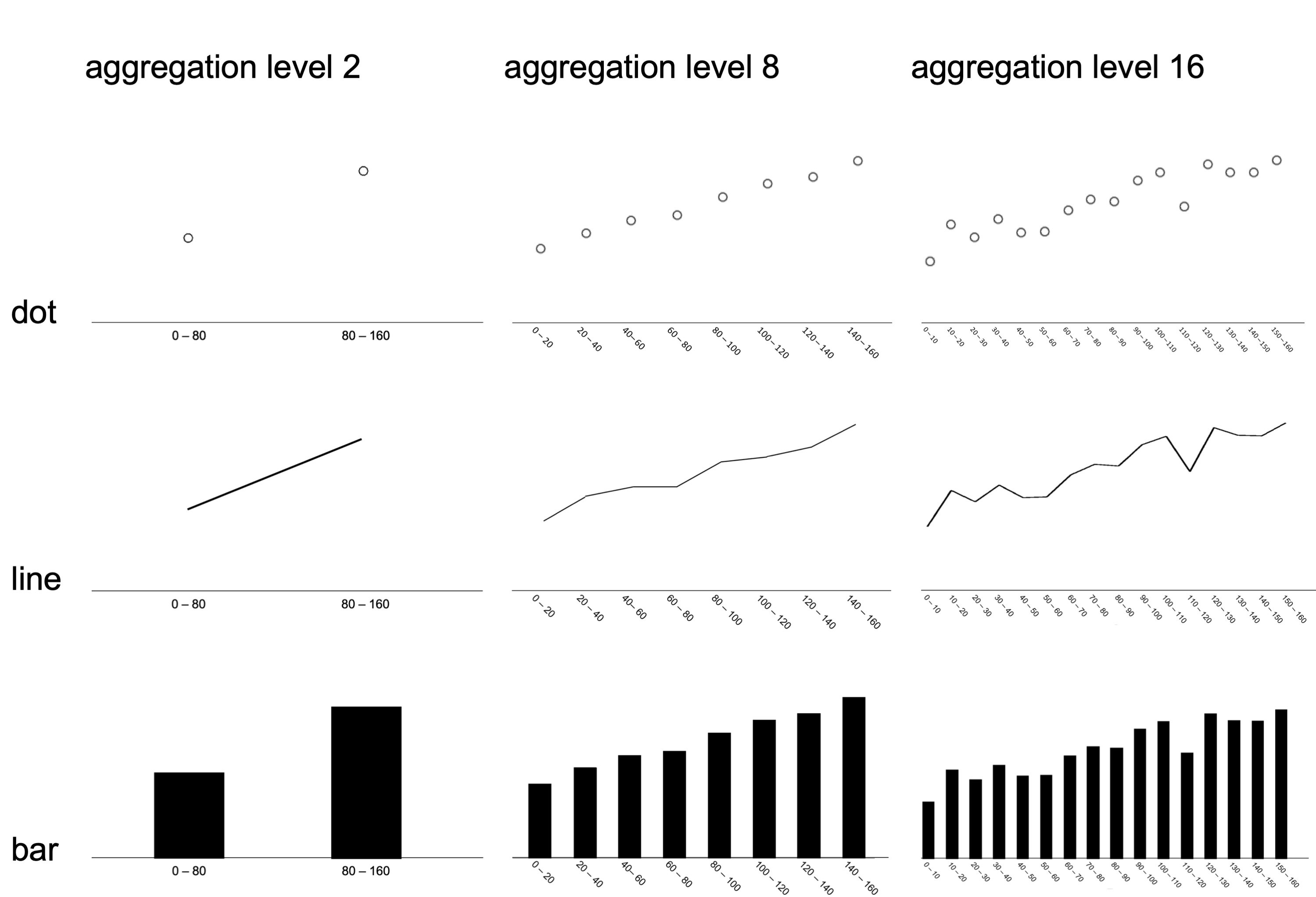}
 \caption{Three aggregation levels tested in Experiment 3 for bar, line and dot type encoding marks.}
 \label{exp3DesignAggregationLevel}
\end{figure}

\subsection{Design}
Because visualization context (i.e., what specific pair of variables was shown) did not influence the effect of visualization design on perceived causality, we omitted context from the visualizations in Experiment 2. Instead of presenting the data in four scenarios with varying plausibility, we stripped the variable names (e.g., "GPA") and replaced with abstract variable labels (e.g., "X","Y"). We operationalized the amount of aggregation as the number of bins the data is sorted in. The bar graph used in Experiment 1 aggregated the data into two bins. For Experiment 2, we additionally created bar graphs that aggregated the data into eight bins and 16 bins. We created dot plots and line graphs using the same binned data in the bar graphs, but replacing the rectangular bars with circles and lines, as shown in Figure \ref{exp3DesignAggregationLevel}. 
Here, bar graphs depict comparisons of data between two, eight or 16 groups, which fit regular conventions of graphic communication using bar graphs\cite{zacks1999bars}. Line charts are also sometimes aggregated, such as when showing daily, weekly, or monthly estimates.  However, conventional scatter plots typically illustrate each dot as an individual data value \cite{sarikayascatter2018}, making our scatterplot stimuli less realistic but useful for the sake of a controlled comparison.  

\begin{figure}[htb]
 \centering
 \includegraphics[width=3.5in]{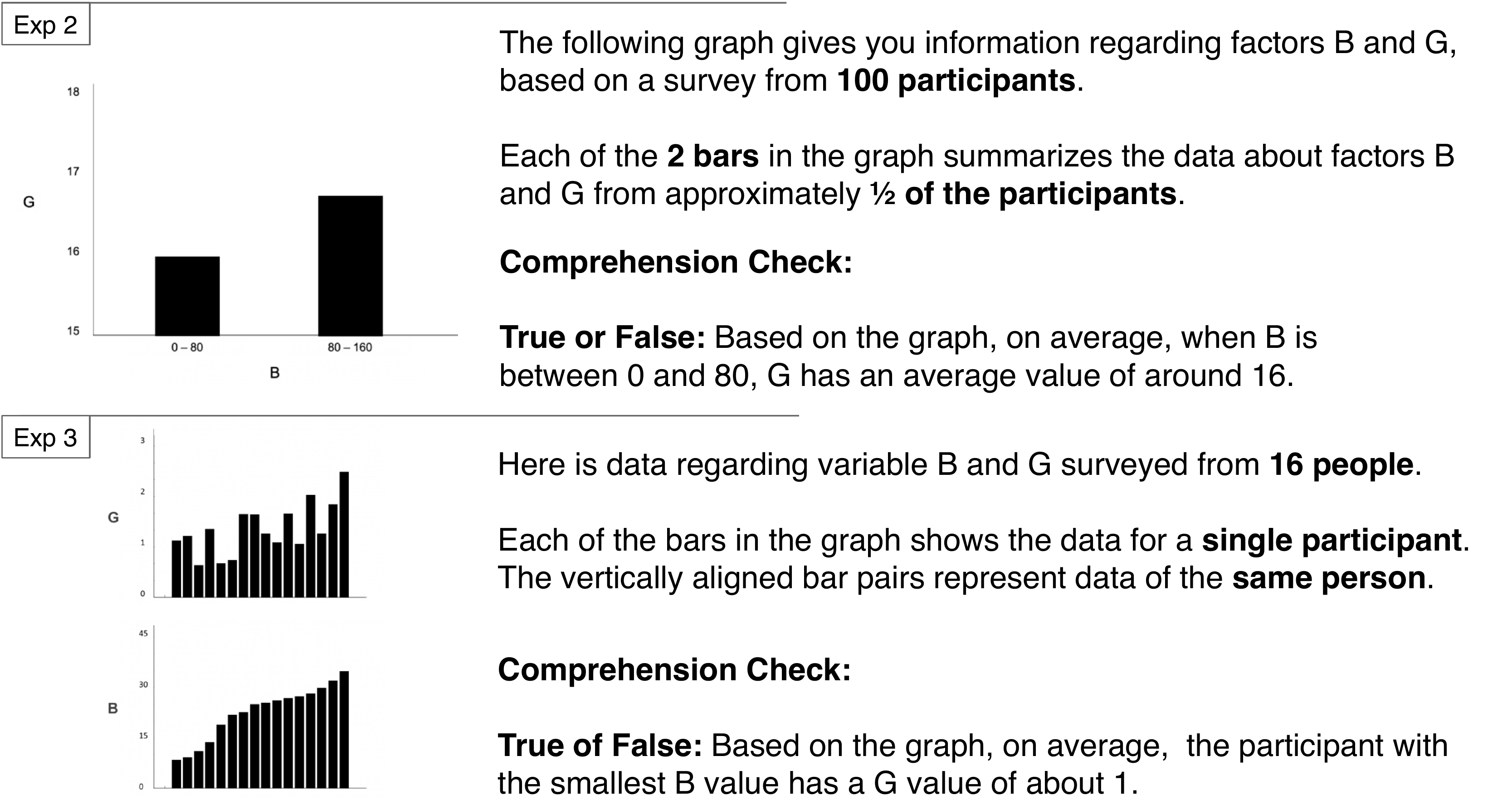}
 \caption{Snapshots from Experiment 2 (left) and Experiment 3 (right).}
 \label{exp3exp4snapshotofexp}
\end{figure}

We explicitly told the participants that the visualized data were generated by summarizing and binning data as they viewed the visualizations, as shown in the left figure in Figure \ref{exp3exp4snapshotofexp}. To ensure the participants understood the plotted data, we created instructions with examples for participants to read through (see supplementary for the example). We asked each participant six graph comprehension questions on the specific visualizations we examined for the experiment, to confirm that participants understood the visualizations, as shown in Figure \ref{exp3exp4snapshotofexp}. Similar to Experiment 1, participants  who failed the comprehension checks were excluded from analysis as they did not appear to have understood the data (the full experiment and data are available as supplementary materials). Participants completed the judgment task by rating how much they agreed with correlation and causation statements, similar to Experiment 1, but we excluded the generative task as the variables were abstract. 

The independent variables in this experiment are visual encoding marks, which can be rectangular bars, lines or dots, and aggregation level, which can be two, eight or 16. The dependent variables are correlation ratings and causation ratings, similar to Experiment 1. We used a $3\times3$ Graeco Latin Square design crossing visualization design and aggregation groups, similar to that in Experiment 1, which crossed visualization design and context. Each participant saw three visualizations --- bar graph, line graph and dot plot, one of which aggregated into two groups, one into eight groups and other into 16 groups. We recruited 129 participants for Experiment 2 using the same method and exclusion criteria.

\begin{figure}[htb]
 \centering
 \includegraphics[width=8.7cm]{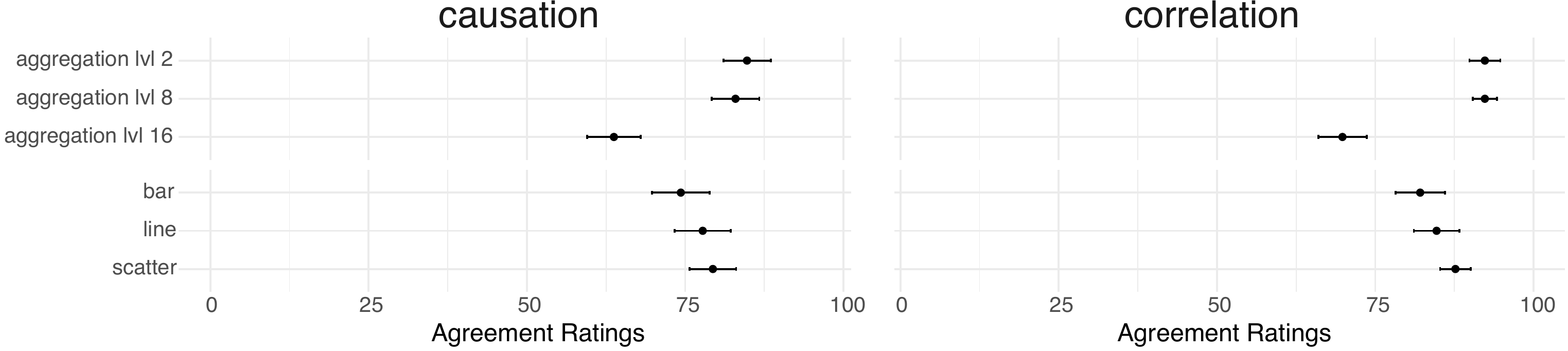}
 \caption{Main effect of aggregation levels (top) and visual encoding types (bottom) on correlation and causation ratings in Experiment 2.}
 \label{ChangethisToo}
\end{figure}

\subsection{Causation Judgment Results}
We used a similar mixed-effect linear model from Experiment 1 to fit the causation ratings with fixed effects of visual encoding marks, aggregation level, an interaction between encoding marks and aggregation level, trial order and demographic information (age, gender, education and political orientation), and a random intercept term accounting for individual differences as random effects. 

The regression model indicated a relatively small main effect of visual encoding marks ($\chi^2$=5.97, $\eta_{partial}^{2}$=0.020,$p$=0.050), such that aggregated dot plots had the highest causality ratings ($M$=79.38, $CI_{95\%}$=[75.67, 83.09]), followed by line encodings ($M$=77.78, $CI_{95\%}$=[73.29, 82.26]), and rectangular bar encodings had the lowest causality ratings ($M$=74.32, $CI_{95\%}$=[69.73, 78.90]), as shown in Figure \ref{ChangethisToo} (top).

There is relatively large main effect of aggregation level, such that visualizations with the more data aggregation were perceived as more causal ($\chi^2$=117.05,$\eta_{partial}^{2}$=0.29,$p$<0.001). Visualizations with aggregation level two, the most aggregation which binned data into two groups, had the highest average causality ratings ($M$=84.76, $CI_{95\%}$=[81.00, 88.55]), followed by visualizations with aggregation level eight ($M$=82.95, $CI_{95\%}$=[79.16, 86.75], and visualization with the least aggregation, which binned data into sixteen groups, had the lowest average causality ratings ($M$=63.74, $CI_{95\%}$=[59.46, 68.03]), as shown in Figure \ref{ChangethisToo} (bottom).

There is an interaction effect between visual encoding marks and aggregation level ($\chi^2$=28.10,$\eta_{partial}^{2}$=0.089,$p$<0.01) on perceived causality, as shown in Figure \ref{fig:allResults}.
For dot encodings, perceived causality did not differ significantly between aggregation level two ($M$=87.19, $CI_{95\%}$=82.54, 91.84]), aggregation level eight ($M$=74.53, $CI_{95\%}$=[66.51,82.56]) and aggregation level 16 ($M$=76.42, $CI_{95\%}$=[70.42,82.41]). For line encodings, perceived causality significantly decreased as the number bins increased, such that aggregation level two ($M$=94.37, $CI_{95\%}$=[91.76,96.98]) was perceived the most causal, followed by aggregation level eight ($M$=84.91, $CI_{95\%}$=[78.55,91.26]), and aggregation level 16 was perceived the least causal ($M$=54.05, $CI_{95\%}$=[46.43,61.67]). For bar encodings, aggregation level eight was perceived as being the most causal ($M$=89.42, $CI_{95\%}$=[84.85,93.98]), followed by aggregation level two ($M$=72.77, $CI_{95\%}$=[63.62, 81.92]), and aggregation level 16 the least causal ($M$=60.77, $CI_{95\%}$=[53.33, 68.20]).

There is a negligible effect of the order the visualizations were presented ($\chi^2$=0.14,$\eta_{partial}^{2}$=0.002, $p$=0.71) as well as participant age, political orientation, gender and education.

\subsection{Comparing Experiment 1 and Experiment 2 Bars}
\label{barCompare}
Experiment 1 seemed to indicate that bar graphs conveyed a greater impression of causation than other representations, Experiment 2 suggests that this impression is due to an interaction between the visual encoding marks and aggregation level. Comparing the causation ratings of bar graphs in Experiment 2 with that in Experiment 1, as shown marked in red in Figure \ref{fig:allResults}, we see that although participants gave lower causation ratings for bar encodings overall, if we only compare the aggregation level two bar condition from Experiment 2 with the bar condition in Experiment 1 (which is an aggregation level two bar graph with context), the two results match ($p = 0.47$), suggesting that bar graphs with two bars may be an interesting case study, see section \ref{lim}. Examining participant quotes for the Experiment 1 in Section \ref{barExperiments} (Causal Conclusions), one explanation may be that many participants associate aggregation level 2 bar graphs with controlled experiments, which can be a valid way to establish causal relationships.

\subsection{Correlation Judgment Results}
We used the same mixed-effect linear model to fit the correlation ratings. The model indicated a relatively small main effect of visual encoding marks ($\chi^2$=9.93,$\eta_{partial}^{2}$=0.03,$p$<0.01), such that aggregated dot plots had the highest correlation ratings ($M$=87.67, $CI_{95\%}$=[85.23, 90.11]), followed by line encodings ($M$=84.69, $CI_{95\%}$=[81.06, 88.32]), and rectangular bar encodings had the lowest ratings ($M$=82.10, $CI_{95\%}$=[78.17, 86.03]), as shown in \ref{ChangethisToo}.

There is a relatively large main effect of aggregation level, such that visualizations with more data aggregation were perceived as more correlational ($\chi^2$=212.31,$\eta_{partial}^{2}$=0.40,$p$<0.001). Visualizations with aggregation level two, the most aggregation which binned data into two groups, had the highest average correlation ratings ($M$=92.32, $CI_{95\%}$=[89.85, 94.79]), followed by visualizations with aggregation level eight ($M$=92.31, $CI_{95\%}$=[90.39, 94.25], and visualization with the least aggregation, which binned data into 16 groups, had the lowest average ratings ($M$=69.82, $CI_{95\%}$=[65.96, 73.68]), as shown in \ref{ChangethisToo}.

There is a medium interaction effect between visual encoding marks and aggregation level ($\chi^2$=30.32,$\eta_{partial}^{2}$=0.088,$p$<0.001) on perceived correlation, as shown in Figure \ref{fig:allResults}. 
For dot encodings, perceived correlation did not differ significantly between aggregation level two ($M$=91.77, $CI_{95\%}$=87.88, 95.66]), aggregation level eight ($M$=88.28, $CI_{95\%}$=[83.49,93.06]) and aggregation level 16 ($M$=82.95, $CI_{95\%}$=[79.12,86.79]). For line encodings, perceived correlations significantly decreased as the number bins increased, such that aggregation level two ($M$=96.42, $CI_{95\%}$=94.49,98.35]) was perceived to be the most correlational, followed by aggregation level eight ($M$=93.37, $CI_{95\%}$=[91.03,95.72]), and aggregation level 16 was perceived  to be the least correlational ($M$=64.28, $CI_{95\%}$=[56.88,71.68]). For bar encodings, aggregation level eight was perceived to be the most correlational ($M$=95.30, $CI_{95\%}$=[93.18,97.43]), followed by aggregation level two ($M$=88.77, $CI_{95\%}$=[82.74, 94.80]), and aggregation level 16 the least correlational ($M$=62.23, $CI_{95\%}$=[55.39, 69.07]).

There is a relatively small effect of the order the visualizations were presented ($\chi^2$=10.65,$\eta_{partial}^{2}$=0.022, $p$=0.001), indicating a learning effect, which is reasonable given the novelty of the visualization designs. There was negligible effect of age and gender, but a relatively small effect of political orientation ($\chi^2$=1.85,$\eta_{partial}^{2}$=0.013, $p$=0.17), such that more liberal participants gave higher correlation ratings overall, and education ($\chi^2$=3 .5,$\eta_{partial}^{2}$=0.019, $p$=0.84), such that participants with higher levels of education gave higher correlation ratings.

\subsection{Discussion of Experiment 2}
Bar visual encoding marks received the lowest causal ratings, followed by line, and dot encodings received the highest causal ratings. These ratings could be further increased or decreased by the amount of data aggregation, such that decreased aggregation (increasing the number of bins) decreased perceived causality, and increased aggregation increased perceived causality in data. However, the visualizations in this experiment all aggregated data, even at the smallest aggregation level (with 16 bins). In order to isolate the effect of visualization encoding, we test how visual encoding marks influence perceived causality when \textit{no} data is aggregated in Experiment 3.

\section{Experiment 3 Effect of Encoding}
The bar graphs and line graphs examined in our first two experiments aggregated data. Experiment 1 showed aggregated bars binned into two groups and a continuous line, which essentially aggregated across all levels. Experiment 2 used aggregated plots which are not commonly seen, because scatter plots and to some extent line charts don't typically depict binned data, as least as often as bar charts do. Scatter plots, for example, usually show non-aggregated raw data. One familiar instance where data is naturally dis-aggregated is a nominal list, which usually shows ranking data, such as \cite{gratzl2013lineup}. 

\subsection{Design and Procedure}
\begin{figure}[htb]
 \centering
 \includegraphics[width=8cm]{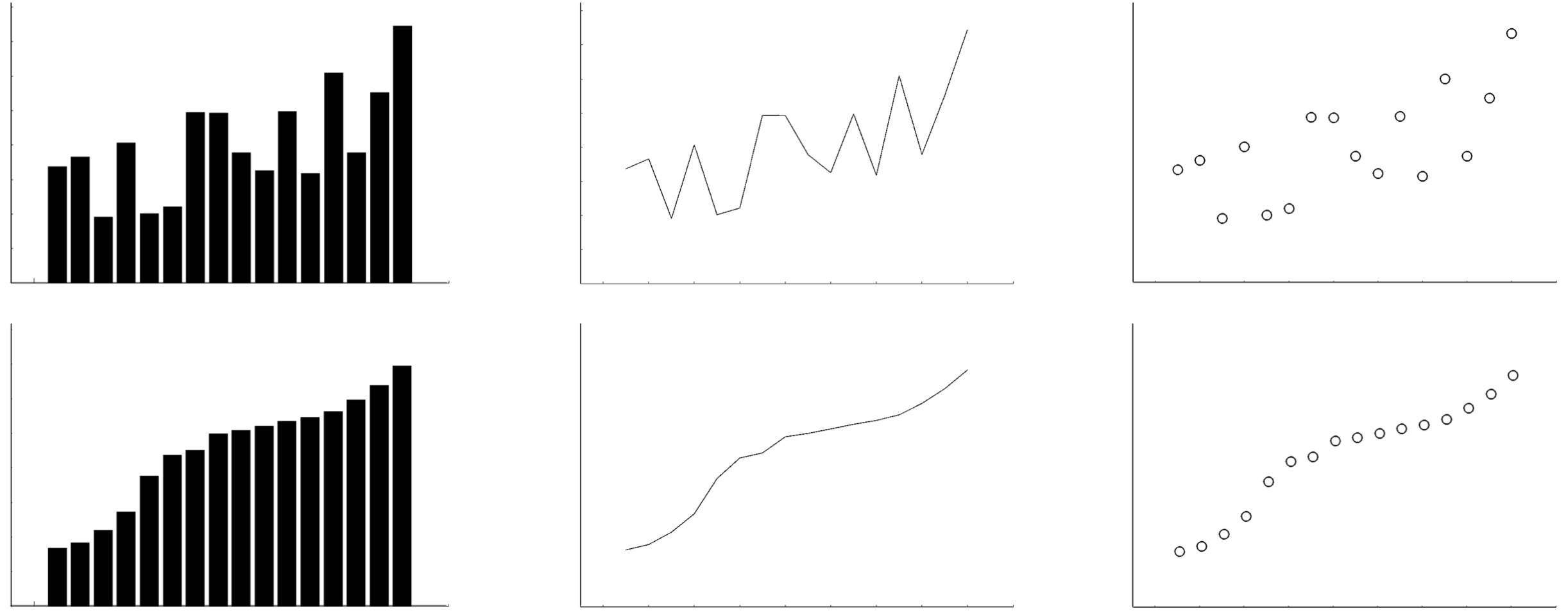}
 \caption{Non-aggregated data visualized with bars, lines and dots.}
 \label{exp4_design}
\end{figure}

We created modified bar graphs, line graphs and scatter plots to present non-aggregated data, as shown in Figure \ref{exp4_design}. This modification aims to parallel the non-aggregated way that scatter plots present data in bar and line charts. For each graph, the x-axis shows the index of each data point. This is a nominal dimension in which order is typically not meaningful, such as an index assigned to each unique name of a person or university. Each of the two graphs shows the value of one variable associated with the index, and the vertically aligned bar pairs represent the variable values associated with the same index. One of the variables was sorted in increasing value to mimic the x-axis and the other is left unsorted mimicking the y-axis in a scatter plot. We made the same modification to line graphs and scatter plots, as shown in Figure \ref{exp4_design}.

Similar to Experiment 2, the visualizations created for this experiment are not conventional and therefore may seem unintuitive to some viewers (although we do sometimes see them in the real world, as shown in the left column of Figure \ref{fig:barMedia}). To ensure the participants in this experiment understood the plotted data, we created instructions with examples for participants to read through (see supplementary for example details). We applied the same exclusion criteria as those in Experiment 2.

In this within-subject design, every participant viewed all three visualization designs in different order, counterbalanced with different axis values labels. An omnibus power analysis, based on pilot effect sizes, suggested a target sample of 62 would yield enough power to detect an overall difference between visualization designs. We collected data following the same data collection and exclusion method as the previous experiments. 


\subsection{Visual Mark Encoding Types}
As shown in Figure \ref{fig:allResults}, a mixed-model linear regression model predicting perceived causality using visual encoding type, trial order and demographic information as fixed effects and individual participants as random effects showed an effect of visual encoding types ($\chi^2$=15.44,$\eta_{partial}^{2}$=0.10,$p$<0.01), such that dot encodings were perceived to be the most causal ($M$=55.49, $CI_{95\%}$=[49.62, 61.36]), closely followed by line encodings ($M$=52.02, $CI_{95\%}$=[46.19, 57.84]) and bar encodings the least causal ($M$=43.21, $CI_{95\%}$=[37.35, 49.07]). There is a relatively small effect of order ($\chi^2$=2.58,$\eta_{partial}^{2}$=0.019) suggesting that participants showed comparatively small learning effects towards the potentially unfamiliar non-aggregated visualizations, age ($\chi^2$=3.43,$\eta_{partial}^{2}$=0.014), such that older participants rated causation less on average, and education ($\chi^2$=4.84,$\eta_{partial}^{2}$=0.035), such that participants with higher levels of education gave higher causation ratings. 

A mixed-model linear regression model predicting perceived correlation using the same fixed effects and random effects showed an effect of visual encoding types ($\chi^2$=15.17,$\eta_{partial}^{2}$=0.10,$p$<0.01), such that dot encodings were perceived to be the most correlational ($M$=60.10, $CI_{95\%}$=[53.86, 66.33]), closely followed by line encodings ($M$=56.27, $CI_{95\%}$=[50.48, 62.06]) and bar encodings the least correlational ($M$=47.86, $CI_{95\%}$=[41.71, 54.00]). There is a relatively small effect of order ($\chi^2$=7.68,$\eta_{partial}^{2}$=0.055) suggesting a relatively small learning effect, and negligible effects of age, gender, political orientation and education.

\subsection{Aggregated and Non-Aggregated Data}
We did a post-hoc between-subject comparison using a mixed-effect linear model comparing the non-aggregated visualization causality ratings in Experiment 3 to the ratings of the visualization with aggregation level 16 in Experiment 2, since both conditions showed 16 data values (16 pairs of values in Experiment 3), differing only in data manipulation -- whether the data was explicitly stated to be aggregated or not. We found a relatively large effect of data manipulation ($\chi^2$=93.38,$\eta_{partial}^{2}$=0.17,$p$<0.001) such that visualizations that aggregated data (Experiment 3, $M$=50.24, $CI_{95\%}$=[46.84, 53.64]) were perceived to be more causal than visualizations that did not (Experiment 2, $M$=77.16, $CI_{95\%}$=[74.70, 79.62]).

\section{General Discussion}
Overall, the choices authors make between visual encoding marks and the amount of data aggregation likely contribute to perceived causality in data. Although our results from Experiment 1 suggest that bar charts were perceived as most likely to be causal, controlling for the amount of data aggregation in Experiment 2 and Experiment 3 suggested that the level of aggregation was the driving factor of higher perceived causality in bar graphs. We also found an effect of visual encoding marks such that bars were perceived to be less causal than line and dot encodings. However, as discussed in section {\ref{barCompare}}, two-bar bar graphs seemed to be a special case where participants consistently perceived the relationship it depicted to be highly causal.

\section{Limitations and Future Directions}
\label{lim}
As an initial investigation of how causality associated with data visualization designs, we feel that it is too early to provide concrete design guidelines to mitigate unwarranted perception of causality in visualized data. We discuss several limitations of the present study and suggest a path forward for future experiments to further understand how visualization design choices impact causality interpretations.

\textbf{Special Case of Two-Bar Bar Graphs: } We suspect there to be something special about two-bar bar graphs that particularly invite causal interpretations, but the present experiments do not confirm the underlying reasons why. Some participant responses suggested that two-bar bar graphs could be  associated with controlled experiments. Future research could confirm whether some inferences are associated with certain visualization types, such as bar graphs with controlled experiments or line graphs with functional relationship between two variables (e.g., $y = f(x)$).

\textbf{Aggregation in Context:} We found no significant effect of context in Experiment 1, and no significant difference between causation ratings of the two-bar bar graph from Experiment 1 (with context) with that from Experiment 2 (no context). Since Experiment 2 and 3 tested abstract variable pairings (e.g., 'G and B'), future work can systematically test how aggregation level might elicit different causal interpretations within the types of concrete context used in Experiment 1.

\textbf{Complex Visualizations:} The present study relied on simple and common data displays, but future work could test more complex displays like dashboards with multiple displays. Some of our studies also relied on displays that were free of context (abstract variable names), and future work should confirm that the results extrapolate to visualizations embedded in context or with explanatory text.

\textbf{Visual Encoding Marks: } We suspect that line encodings were most likely to be associated with causality because line encodings are likely associated with continuous trends in data, which could have made the line encoding marks appear more correlational, and thus more causal. Dot encoding types, although conventionally associated with non-aggregated raw data, still depict apparent trends in data as participants could mentally draw lines connecting each points. Bar encodings, in contrast, are visually vertically asymmetrical, with the area below the mean filled and the area above unfilled. In light of previous work on bar graphs showing that this vertical asymmetry invites perceptual and cognitive biases \cite{correll2014error, newman2012bar}, we speculate the vertical asymmetry made the trends in bar encoding visuals more difficult to see than trends in line and dot visuals, thus appearing less correlational, and therefore less causally perceived. Future research should empirically test our hypothesis to further understand visual reasons \textit{why} bar encodings were perceived less causal than line and dot encodings. 

\textbf{Other Data Sets: }We used the same data set to create the visualization designs in these experiments, which means the correlation depicted was always an upward trend. We purposefully chose this positive trend to avoid common reasoning errors such as misinterpreting negative correlations to be smaller than the actual correlation \cite{huff1993lie}. While the goal of this experiment is to investigate whether visualization design can elicit varying degrees of perceived causality in data, further research should investigate the impact of the strength and direction of the correlation. 

\textbf{Improving Taxonomy for Generative Task Evaluation: } Our qualitative characterization of verbal responses could be improved. We encountered several instances of ambiguous language, such as "\textit{there is some sort of relationship between A and B}," which made it difficult for researchers to decide whether the participants meant a correlation or a causal relation. Some participants used template phrases such as "\textit{correlation is not causation}" and "\textit{A is correlated with B}" to describe relations in data, but we lacked ways of evaluating whether they \textit{actually} read a causal relation from the data or not. 

\textbf{Statement Choices: }The present experiment only presented one type of correlation and causation statement for participants to rate their level of agreement. We purposefully avoided directly using words like `correlation' and `causation' to better evaluate participants' interpretation of the \textit{visualized} data instead triggering knee-jerk reactions to the \textit{words} `correlation' and `causation.' Future iterations of the experiment should test how participants would react differently to other types of statements, such as direct causal statements and non-counter-factual statements. We also did not randomize the question order such that participants always responded to the generative task first, and then the judgment task rating correlation statements followed by causation statements. Participants could be using the correlation statements as a `baseline' to their causation statement ratings. Future research can also investigate the extent to which changing question order would influence correlation and causation ratings. 

\textbf{Alternative Ways to Prevent Causal Interpretations: } Our work took an initial step toward showing that visualizations can be designed to mitigate misinterpretation of correlation and causation. Future experiments could investigate how other techniques, such as verbal annotation on the visualization, could reinforce better interpretation of correlation and causation in addition to visualization designs, potentially contributing to data journalism and education.

\bibliographystyle{abbrv}
\bibliography{reference}
\end{document}